\documentclass[preprint]{aastex}

\shorttitle{3-D MHD pulsed jets}
\shortauthors{Cerqueira \& de Gouveia Dal Pino}

\begin{document}

\title{Three-dimensional MHD simulations of radiatively cooling, pulsed jets}

\author{Adriano H. Cerqueira\altaffilmark{1} \& Elisabete M.  de Gouveia
Dal Pino} \affil{Instituto Astron\^{o}mico e Geof\'{\i}sico, Universidade
de S\~{a}o Paulo, \\ Av. Miguel St\'efano 4200, (04301-904) S\~{a}o
Paulo - SP, BRAZIL \\ adriano@quark.iagusp.usp.br, dalpino@iagusp.usp.br}

\altaffiltext{1}{Current address: Departamento de Ci\^encias Exatas
e Tecnol\'ogicas, Universidade Estadual de Santa Cruz, Rodovia
Ilh\'eus/Itabuna, km 16, Ilh\'eus, BA, 45650-000, BRAZIL}

\begin{abstract}

We here investigate, by means of fully three-dimensional (3-D)
Smoothed Particle Magnetohydrodynamic (SPMHD) numerical simulations,
the effects of different initial magnetic field configurations on
the evolution of overdense, radiatively cooling, pulsed jets, using
different initial magnetic field topologies: (i) a longitudinal, (ii)
a helical geometry permeating both the jet and the ambient medium, and
(iii) a purely toroidal geometry permeating the jet only. We explore
the effects of different pulsational periods, as well as different
values of the magnetic field strength ($\beta \simeq 0.1- \infty$,
or $B \simeq 260$ $\mu$G-0).  The presence of a helical or a toroidal
field tends to affect more the global characteristics of the fluid than
a longitudinal field. However, the relative differences which have been
previously detected in 2-D simulations involving distinct magnetic field
configurations are diminished in the 3-D flows. While the presence of
toroidal magnetic components can modify the morphology close to the jet
head inhibiting its fragmentation in  the early evolution of the jet, as
previously reported in the literature, the impact of the pulsed-induced
internal knots causes the appearance of a clumpy, complex morphology
at the jet head (as required by the observations of Herbig-Haro jets)
even in the MHD jet models with helical or toroidal configurations. The
detailed structure and emission properties of the internal working
surfaces can be also significantly altered by the presence of magnetic
fields. The increase of the magnetic field strength (decrease of $\beta$)
improves the jet collimation, and amplifies the density (by factors up
to 1.4, and 4) and the H$\alpha$ intensity (by factors up to 4, and 5)
behind the knots of jets with helical field and $\beta \simeq 1-0.1$
(respectively), relative to a non-magnetic jet.  As a consequence,
the  corresponding $I_{{\rm [S~II]}}/I_{{\rm H}\alpha}$ ratio (which is
frequently used to determine the excitation level of HH objects) can be
largely decreased in the MHD models with toroidal components relative
to non-magnetic calculations.  We also find that the helical mode of the
Kelvin-Helmholtz instability can be triggered in MHD models with helical
magnetic fields, causing some wiggling of the jet axis.  No evidence
for the formation of the nose cones that are commonly detected in 2-D
jet simulations with initial toroidal magnetic fields, is found  in the
3-D flows, nor even in the $\beta \simeq 0.1$ case.  The implications
of our results for Herbig-Haro jets are briefly discussed.

\end{abstract}


\section{Introduction}

The bright Herbig-Haro (HH) objects associated with low-mass young stellar
objects, are radiating shock fronts immersed in collimated optical jets
that may extend from a few 1000 AU (e.g., the microjets in the Orion
Nebula; Bally, O'Dell \& McCaughrean 2000), to very large parsec-length
scales, like the $giant$ HH jets (Heathcote et al. 1998; Bally \& Devine
1994; see also Reipurth 1999 for a catalogue of such objects).

Currently, the most accepted model for the production of the discrete HH
knots is by time variability in the driving sources of the jets. Their
bow shock morphology, high spatial velocity, and symmetric pattern with
respect to the star (e.g., Reipurth 1989a, 1989b; B\"urke, Mundt \&
Ray 1988; Zinnecker, McCaughrean \& Rayner, 1997, 1998) all indicate
that the HH shocks generally arise through the steepening of velocity
fluctuations in the underlying, supersonic outflow. Strong support
for this conjecture has been given by theoretical studies which have
confirmed that traveling shocks created in this manner reproduce the
essential properties of the observed knots (e.g., Raga et al. 1990; Raga
\& Kofman 1992; Kofman \& Raga 1992; Stone \& Norman 1993; de Gouveia
Dal Pino \& Benz 1994; V\"olker et al. 1999; de Gouveia Dal Pino 2001).
\footnote{We note that although pinch-modes of the  Kelvin-Helmholtz (K-H)
were originally suggested as the source of the knotty pattern in HH jets,
it has been found that these modes have their strength highly reduced in
the presence of radiative cooling (e.g., Blondin, Fryxell and K\"onigl
1990; de Gouveia Dal Pino \& Benz 1993). Besides, they typically lead
to a too fast periodicity to match the observations (e.g., Stone, Xu,
and Hardee 1997). Nevertheless, nonaxisymmetric K-H modes appear to be
responsible for the gentle wandering or wiggling observed in many jets
(e.g., de Gouveia Dal Pino, Birkinshaw, \& Benz 1996; Stone, Xu, \&
Hardee 1997).}

Magnetic fields seem to play a fundamental role in several phases of
the jet launching and propagation processes. They seem to be relevant
to: (i) operate locally in an accretion disk, in order to regulate the
angular momentum process (Balbus \& Hawley 1998) and the accretion rate;
(ii) launch outflows through magneto-centrifugal forces from the disk
itself (e.g., Blandford \& Payne 1982; Kudoh, Matumoto \& Shibata 1998;
K\"onigl \& Pudritz 2000), or  from the disk-star boundary region
(the so-called X-point; e.g., Shu et al. 2000); and (iii) collimate
the outflow once launched (e.g., Ouyed, Pudritz \& Stone 1997). On the
larger scales, since  the outflowing material does not seem to be highly
resistive (e.g., Frank et al.  1999),  the magnetic field may also play
a relevant role on its dynamics and evolution. Previously, Morse et
al. (1992) have inferred an upper limit for the magnetic field in the
ambient medium upstream the HH34 bow shock of about $10-20 \times 10^{-6} $
G. More recently, Ray et al. (1997), based on polarization measurements,
have detected magnetic field strengths of the order 1 G at a distance
of few tens of AU from T Tau S. Using magnetic flux conservation, this
latter result would give magnetic fields of the order of $\sim 10^{-3}
$ G for a toroidal field configuration, and $\sim 10^{-6} $ G for a
longitudinal field at distances $\sim 0.1$ pc. For typical jet parameters,
this would imply a plasma parameter $\beta \, = \, p_{j,gas}/(B^2/8 \pi)
\, \simeq \, 10^{-3}$ to $10^{3}$ at those distances. Although Ray et
al. have concluded that the high values measured for B at $\sim 10 $
AU probably come from regions of strong magnetic field amplification
(such as behind shocks), the figures above suggest that magnetic fields
must play some role in the overall jet dynamics and propagation even
on the outer scales where local amplification (particularly behind
the leading and the internal bow shocks) is expected to occur (e.g.,
Cerqueira, de Gouveia Dal Pino \& Herant 1997; hereafter, Paper I).

Lately, there has been increasing focusing on MHD studies of the
propagation of overdense, radiative cooling jets in an attempt to
look for possible  signatures of magnetic fields on the large scales
of the HH outflows (see, e.g., de Gouveia Dal Pino \& Cerqueira 2001
for a review). Frank et al. (1997, 1998), for example, in  axisymmetric
studies of steady-state jets with toroidal magnetic fields (and $\beta =
8\pi p_{th}/B^2 < 1$), have reported the formation of pinch modes driven
by magnetic tension, and the appearance of a magnetically confined nose
cone structure of compressed gas between the bow shock and the Mach disk
at the jet head.

In  previous investigations (de Gouveia Dal Pino \& Cerqueira 1996; Paper
I; Cerqueira \& de Gouveia Dal Pino 1999, hereafter Paper II), we have
employed a three-dimensional Smoothed Particle Hydrodynamic (SPH) code
that has been implemented to compute the effects of magnetic fields to
investigate steady-state jets. Assuming different initial magnetic field
configurations (in approximate equipartition with the gas) permeating
both the jet and the ambient medium, we have found that there is an
increase in the jet collimation in comparison to purely hydrodynamic
calculations, which is caused by the amplification and reorientation of
the magnetic fields behind the shocks (particularly in the presence of
a helical field). Further, we have found that the presence of a  helical
magnetic field may inhibit the fragmentation of the dense shell formed by
cooling and Rayleigh-Taylor instability at the head of the jet (Paper I),
while a longitudinal field retains the fragmentation. This result could be
an indication that the latter configuration could dominate near the jet
head as the observed jets have a clumpy structure there. We have also
evidenced from the simulations the development of the MHD K-H modes,
but the weakness of the pinch modes makes it doubtful that they could
play an important  role in the production of the bright knots (Paper
II). Where there is overlap, these results have been confirmed by recent
axisymmetric  calculations (e.g., Gardiner et al. 2000; Stone \& Hardee
2000; O'Sullivan \& Ray 2000). In particular, Gardiner et al. (2000),
assuming longitudinal magnetic fields (with $\beta = 0.1 \, - \,10^{7}$),
have found that the global characteristics of the flow are not strongly
affected by the $\bf{B}$-field strength, and a predominantly axial field
tends to increase the collimation and order, and inhibit instabilities
in the flow, but the presence of internal pulses increases the likelihood
of magnetic reconnection.

The 2-D calculations of steady-state and pulsed MHD jets of Stone \&
Hardee (2000; hereafter, SH) propagating into unmagnetized ambient
media have also shown that the magnetic fields affect the fragmentation
of the shell at the jet head, and a strong toroidal field peaking
near the jet surface develops a nose cone which should be unstable in
three-dimensions. They have also found that the radial hoop stresses due
to the toroidal field confine the shocked jet material in the internal
pulses resulting in higher densities in the pulses which are strongly
peaked towards the jet axis in comparison to purely hydrodynamic
calculations.  A similar line  of numerical studies has also been
conducted by O'Sullivan \& Ray (2000;  hereafter OR). Most of their
results essentially confirm previous hydrodynamic (e.g., Blondin, Fryxell,
\& K\"onigl 1990; de Gouveia Dal Pino \& Benz 1993, hereafter GB93; Stone
\& Norman 1994) and the MHD results above. However, unlike these works,
they have assumed jets with initially lower densities  with respect to
the ambient medium (corresponding to a density ratio $\eta$ = 1), and
which are highly over-pressurized with respect to the ambient medium.
This causes the development of more complex cocoons surrounding the
beams and the formation of crossing shocks which help to  refocus the
beam  and the internal pulses. In particular, they find that the  hoop
stresses associated with toroidal fields can cause the $disruption$
of the internal knots, even for $\beta \simeq $1.

In a recent study, we have presented the first results of 3-D MHD
simulations of the early evolution of pulsed jets (Cerqueira \& de Gouveia
Dal Pino 2001, hereafter, Paper III).  In the present paper, we attempt
to extend these previous investigations by exploring in three-dimensions
the role played by  three different initial magnetic field configurations
on the evolution and emission structure of radiatively cooling, pulsed
jets, considering a more extensive range of parameters.  As before (Paper
I, II and III), together with a baseline non-magnetic calculation, we
employ a modified version of our 3-D SPH magnetized code and consider a
longitudinal and a helical field topology (in approximate equipartition
with the gas), both permeating the jet and the ambient medium. Also, in
order to make a closer comparison with previous studies, a third geometry
involving a purely toroidal field permeating the jet is considered.

The paper is planned as follows. In \S 2, we will briefly outline the
numerical technique and setup. In \S 3, we present the results of the
simulations for pulsed jets; in \S 4, we compare the results with previous
2-D MHD calculation and discuss some observational implications of them,
and present our final remarks and conclusions.

\section{Numerical method}

We solve the equations that describe the evolution of a fluid using
the Smoothed Particle Hydrodynamic (SPH) technique (e.g.; Benz 1990;
Monaghan 1999). SPH is a Lagrangean scheme that makes use of particles
to follow fluid parcels, and the evaluation of a given physical quantity
is given by an interpolation scheme, using for this a kernel function
($W$). The kernel has an adaptative smoothing length ($h$) value,
which defines the space volume of the SPH particles.  Like any other
physical quantities, magnetic fields can also be handled with SPH. In
order to take into account the presence of the magnetic field, we have
modified a pure HD, SPH code.  In Paper II, we give the details of such
modification, as well as the results of some tests in the MHD limit. The
application of the SPH technique to pure hydrodynamic jet simulations
can be seen in several papers (e.g., GB93; de Gouveia Dal Pino \& Benz
1994; Chernin et al. 1994; de Gouveia Dal Pino \& Birkinshaw 1996; de
Gouveia Dal Pino, Birkinshaw, \& Benz 1996; de Gouveia Dal Pino 1999,
2001), so that our code has been continuously tested in both HD and
MHD regimes. Such an MHD, SPH based code is often referred as ``SPMHD"
(e.g., Stellingwerf \& Peterkin 1990; Meglicki 1994).

The solved system of MHD equations, in its ideal approximation, is
given by:

$$\frac{d{\rho}}{dt}=-\rho{\bf {\nabla}}\cdot{\bf v} \eqno(1a)$$

$$\frac{d{\bf v}}{dt}=-\frac{{\bf
{\nabla}}p}{\rho}+\frac{1}{4{\pi}{\rho}}( {\bf {\nabla}}\times{\bf
B})\times{\bf B} \eqno(1b)$$

$$\frac{du}{dt}=-\frac{p}{\rho}({\bf {\nabla}}\cdot{\bf v}) - {\cal L}
\eqno(1c)$$

$$\frac{d{\bf B}}{dt}=-{\bf B}({\bf {\nabla}}\cdot{\bf v})+ ({\bf
B}\cdot{\bf {\nabla}}){\bf v} \eqno(1d)$$

\noindent where $\rho$ is the density; ${\bf B}$ is the magnetic field;
$u$ is the specific internal energy and ${\cal L}$ is the radiative
cooling rate. As in our previous works (see Paper II and references
therein), we have adopted the cooling function given by Katz (1989) for
a gas of cosmic abundances cooling from $T \simeq 10^6$K to $\simeq
10^4$K. An ideal equation of state is used to close the above system
of equations:

$$p=(\gamma -1)\rho u \eqno(1e)$$

\noindent with $\gamma=5/3$.

\subsection{Initial and boundary conditions}

Our computational domain is a rectangular box with dimensions $-30 R_j
\le$ x $\le 30 R_j$, and $-10R_j \le$ y,z $\le 10 R_j$, where $R_j$
is the initial jet radius (which is also the code distance unit). The
Cartesian coordinate system has its origin at the  center of the box
and the jet flows through the x-axis. The jet is continuously injected
into the bottom of the box [at ${\bf r}=(-30R_j,0,0)$]. Inside the box,
the SPH particles are initially distributed on a cubic lattice. An
outflow boundary condition is assumed for the boundaries of the box.
The particles are smoothed out by a spherically symmetric kernel function
of width $h$, and the initial values of $h$ were chosen to be $0.4 R_j$
and $0.2R_j$ for the ambient and jet particles, respectively, so that
we have up to 400,000 SPH particles at the beginning of the calculation.

We adopt a sinusoidal velocity profile for the pulsing jet at the inlet:

$$v_o(t)= v_j [1 + A\cdot{\rm sin}({{2 \pi}\over P} t)] \eqno(2)$$

\noindent where $v_j$ is the mean jet speed, and $A$ and $P$ are the
amplitude and the period of the velocity oscillations, respectively.
We have chosen two periods for our models, namely, $P=1$ and $P \sim 0.5$
(in code units, where the unit of time in our calculations corresponds
to the ratio $t_d = R_j/c_a \sim 38$ years, for the set of parameters
adopted here; see below). The number density in the flow at injection,
$\rho_j$, is assumed to be constant, which implies that $\dot{m}=\rho_j
\cdot v_j(t)$ is not constant. These models are equivalent to the
constant density injection models recently investigated analytically
by Cant\'o et al. (2000).

We consider different initial magnetic field profiles.  As in papers I,
II and III, we adopt 1) an inilial constant longitudinal magnetic field
[${\bf{B}}=(B_0,0,0)$], both inside and outside the jet beam; and 2)
a helical force-free field that also extends to the ambient medium
which is described by the following equations (see Todo et al. 1993,
and Fig. 1 of Paper II):

$$B_r=0 \eqno(3a)$$

$$B_{{\phi}}(r)=B_0{\bigg[} \frac{0.5Cd r^2}{(r+0.5d)^3}{\bigg]}^{1/2}
\eqno(3b)$$

$$B_x(r)=B_0{\bigg[}1- \frac{Cr^2(r + d )}{(r+0.5 d )^3} {\bigg]}^{1/2}
\eqno(3c)$$

\noindent where $r=\sqrt{y^2+z^2}$ is the radial distance from the
jet axis and the $C$ and $d$ constants are given by 0.99 and $3R_j$,
respectively. In these equations, $B_0$ is the maximum strength of the
magnetic field and corresponds to the magnitude of the longitudinal
component at the jet axis.  For these configurations, the jet is assumed
to have an initially constant gas pressure ($p_j$) which is in equilibrium
with the ambient gas pressure ($\kappa=p_j/p_a=1$).

The third adopted configuration is a purely toroidal magnetic field
($B_{\phi}$) whose functional form is given by  Fig. 1 and equation (5)
of SH.  Only the jet beam is initially magnetized in this case, and for
the sake of clarification, we here rewrite SH's equation for $B_{\phi}$:

$$
B_\phi (r)=\left\{ \begin{array}{cc}
  B_{\phi ,m}\frac r{r_m} & 0\leq r\leq r_m \\
  B_{\phi ,m}\frac{R_j-r}{R_j-r_m} & r_m\leq r\leq R_j \\
  0 & R_{j}<r \end{array} \right.
\eqno(4)$$

\noindent where $r_m$ is a free parameters that determines the position
where the magnetic field has its maximum intensity (we adopt here $r_m
\sim 0.9 R_j$). Lind et al. (1989), Frank et al. (1997, 1998) and OR,
have also used a similar toroidal profile.  In this case, in order
to ensure initial magnetostatic equilibrium, the jet gas pressure has
a radial profile with a maximum at the jet axis $p_j(0) \approx 2.07
p_j(R_j)$, and $p_j(R_j)=p_a$ at the jet surface (see Figure 1 of SH).

The physical conditions at the jet inlet are parameterized by the
following nondimensional numbers: $\eta=n_j/n_a$, the jet-to-ambient
number density ratio; $M_a=v_j/c_a$, the average jet velocity to the
ambient sound velocity ratio; $\kappa=p_j/p_a$, the jet-to-ambient thermal
pressure ratio; $\beta=8 \pi p_{th}/ B^2$, the thermal-to-magnetic
pressure ratio.  As we will also study jets in which  the magnetic
field is a function of the radial distance $r$, we can define a mean
magnetossonic Mach number as being $\langle M_{ms} \rangle = \int M(r)rdr
/ \int r dr$, where $M(r)=v_j/[c + v_A(r)]^{1/2}$ (e.g., SH), and $ v_A$
is the Alfven speed (see Table 1).

\subsection{ Magnetic Field Reversals}

We should make some remarks on the late evolution of radiatively
cooling magnetized jets containing longitudinal components.  In Paper
II, we have shown examples with the development of magnetic field
reversals at the contact discontinuity between the jet and the cocoon
with intensities up to 5 times their initial magnitude. As stressed in
Paper II, field reversals of the longitudinal component occur in both
purely longitudinal and helical magnetic field configurations, because
the field lines are amplified by compression in the nonparallel shocks
at the jet head, and are enforced to flow backward with the shocked
plasma into the cocoon. In this process, the lines  are reoriented and
sometimes have their polarization reversed.  The increasing strength of
the reversed fields due to shear at the contact discontinuity could lead
to the development of strong pinching regions in the late evolution of
some flows that ultimately could cause jet disruption (see Fig. 16 of
Paper II). Although shear and compression are expected to enhance {\bf
B}, we have speculated in Paper II that very large amplifications of
the reversed fields could be due either to inappropriate computation
of field dissipation under our ideal-MHD treatment, or to numerical
effects.  

In order to test the first of these hypotheses, we have
carried out numerical simulations under a non-ideal MHD approximation,
using different values for the magnetic resistivity. Although a complete
analysis of the results of these tests is out of the scope of the present
paper, we can briefly comment their  consequences to the present study.
We have found that a non-null magnetic resistivity ($\eta_M$) is able
only to postpone the effects above at the jet/cocoon interface [tested
for a broad range of values of  $\eta_M$ from the thermal value (which
is $\propto T^{-3/2}$; e.g., Spitzer 1956), to values many orders of
magnitude higher].  

These findings have suggested a possible numerical
origin for the large field reversals found in the earlier calculations.
Performing then, extremely high resolution SPMHD calculations, using up to
10$^6$ SPH particles and the same conditions as in Fig. 16 of Paper II,
we could eliminate the anomalous amplification of the reversed fields
and the disruption of the beam, while the other features that appeared
in the original lower resolution model were left unaffected both at the
head and along the jet beam.  The anomalous amplification of the fields
was, therefore, being caused essentially by a poorer interpolation
of the physical quantities at the contact discontinuity in the lower
resolution models.  As the extremely high resolution simulations expend
too much computational time and memory without significantly improving
the overall results, we have instead, presently introduced in our code
a numerical $cleaning$ $switch$, in order to eliminate  the anomalous
magnetic fields at the jet/cocoon interface, in calculations involving
logitudinal magnetic fields.  The cleaning switch allows the reduction
of the strength of the magnetic fields at the jet/cocoon interface by
an amount of about 1\% whenever  the reversed $\bf{B}$-field component
exceeds a critical value that was chosen to be  twice as large as
the maximum value of the field  in the head (where $\bf{B}$ regularly
reaches its maximum physical amplification behind the strong shocks).
In average, less than 0.01\% of the SPH particles have their magnetic
field vectors modified by such switch, which is applied in only less
than 1\% of the total number of time steps. The tests have indicated
that the system is stabilized against anomalous field amplification and
jet disruption with the employment  of this switch and, more important,
the results of these models are comparable with those obtained in the
extremely high resolution calculations.

\section{The simulations}

In this section, we present the results of our 3-D, MHD numerical
simulations of radiatively cooling pulsed jets. The physical parameters of
our models are summarized in Table 1. All the parameters are suitable for
the study of protostellar jets, namely: $R_j \sim 2.5\times 10^{15}$ cm,
$M_a \approx 15$, or $v_j \approx 15 \times c_a \approx 250$ km s$^{-1}$;
$\eta=n_j/n_a=5$; $n_a = 200$ cm$^{-1}$ (e.g., Reipurth \& Raga 1999).
We have adopted in most of the simulated models, a maximum intensity
for the magnetic field as given by the equipartition condition: $\beta=8
\pi p_{th}/B^2 \simeq 1$, which implies a maximum intensity of $\approx$
80 $\mu$G.  In some cases, we have adopted 
$\beta \approx 0.1$, which implies a maximum $B \approx 260\mu$G. The
MHD models are all compared with the purely HD counterparts for which
$\beta = \infty$.

\subsection{Effects of different magnetic field configurations}

Figure 1 displays the midiplane density contours (left) and the velocity
field distributions (right) for four supermagnetosonic, radiatively
cooling, pulsed jets with initial $\beta \simeq 1$ after they have
propagated over a distance $\approx 35R_j$, at $t/t_d \simeq 3$. The
top jet is purely hydrodynamical (model HD1 in Table 1); the second jet
(from top to bottom) has an initial constant longitudinal magnetic field
(model ML1 in Table 1); the third jet has an initial helical magnetic
field (MH1), and the bottom jet has an initial toroidal magnetic field
(MT1). The early evolution of these jets was discussed in Paper III.

At the time depicted in Figure 1, the leading working surface at the jet
head is followed by 4 other features. Each new pulse at the left forms
within a time of about $0.2 t_d$. This value agrees with that predicted
by linear analysis of pulsed jets with sinusoidal injection profile (Raga
et al. 1990; Raga \& Cant\'o 1998). Like the leading working surface, each
internal feature consists of a double-shock structure, an upstream reverse
shock that decelerates the high velocity material entering the pulse, and
a downstream forward shock  sweeping up the low velocity material ahead of
the  pulse. Each of these internal working surfaces (IWS) or knots, widens
and broadens as it propagates downstream and squeezes shocked material
sideways into the cocoon.  Since $P \simeq 0.5t_d$, there has been time
for 5 IWSs to form through steepenning of the input sinusoidal profile.
The first of these IWSs has already overtaken the leading working surface
and merged with it. With the impact, it is disrupted and its debris are
partially deposited into the cocoon thus providing a complex fragmented
structure at the head as required by the observations (as is the case,
for example, of HH 1/2; e.g., Jeff Hester, Stapelfeldt \& Scowen 1998).

As previously reported (see Paper III, and de Gouveia Dal Pino \& Cerqueira
2001), Fig. 1 suggests that the overall morphology of the 3-D pulsed jet
is not very much affected by the presence of the different magnetic field
configurations in equipartition with the gas. Compared with the purely HD
jet (top panel), the introduction of the distinct $\bf{B}$-profiles tends,
instead, to alter essentially the detailed structure and the emission
properties (see below) behind the shocks at both the head and internal
knots, particularly when helical or toroidal fields are present. For
example, like in steady-state calculations (see Paper I and II), while
the MHD jet with initial longitudinal $\bf{B}$-field (second panel)
exhibits a fragmented dense shell at the head which is very similar to
that in the pure HD jet, the MHD jets with helical and toroidal fields
present a more collimated head structure due to the action of the tension
forces associated with the toroidal component of the $\bf{B}$-field
(third and fourth panels; see also Paper III).

Several physical quantities along the beam axis of the jets of Fig. 1 with
initial longitudinal (ML1), and toroidal (MT1) fields are displayed in
Figures 2a and 2b, respectively.  As previously reported for the helical
jet of Fig. 1 (see Paper III), we see (top panel, Figs. 2a and b) that
the initial sinusoidal velocity profile impressed on the flow at the
inlet steepens into the familiar sawtooth pattern (e.g., Raga \& Kofman
1992) as the faster material catches up with the slower, upstream gas in
each pulse.  The longitudinal field ($B_{x}$) depicted in Fig. 2a (bottom
panel) is anti-correlated with the sharp density peaks within each IWS
(second panel), where it  attains a minimum value, and amplifies between
them. On the other hand, the toroidal magnetic field ($B_{\phi}$) depicted
in Fig. 2b (bottom panel) sharpens within the knots and rarefies between
them. These results are in agreement with those found in Paper III for
the jet with initial helical geometry (see also Gardiner \& Frank 2000).

Figures 3a and b depict the same HD jet (HD1, Fig. 3a), and the MHD
jet with helical field (MH1, Fig. 3b) of Figure 1, but both more evolved,
i.e., after they have propagated over a distance $\simeq 60R_j$ ($t/t_d
\simeq 5$). We note in both the development of a gentle wandering
or wiggling along the jet  axis which is much more pronounced in the
MHD jet. This is excited by the kink mode of the Kelvin-Helmholtz
(K-H) instability. From the simulation we obtain a wavelength for the
oscillation $\lambda \approx 16R_j$, which is in good agreement
with that derived for the resonant wavelength of the fundamental
kink mode of the K-H instability from the linear theory (e.g., Hardee,
Clarke \& Rosen 1997; Hardee \& Stone 1997, Hardee \&
Norman 1988). This effect could provide a potential explanation for the
wiggling features often detected in HH jets (e.g., HH 111 and HH 46/47;
Reipurth et al. 1997; Heathcote et al. 1996).

Figure 4 displays the temporal evolution of the density of the leading
working surface (left panels) and of the second IWS (right panels) for the
four jets of Figure 1: HD1 (top), ML1 (second panel), MH1 (third panel),
and MT1 (bottom). As reported in previous 3-D calculations of steady-state
jets (e.g., GB93; Paper I and II), the density of the leading working
surface in the HD1 model (top panel, left) shows a temporal variability
that is caused by global thermal instabilities (GTI) of the strong,
high velocity radiative shocks (e.g., Gaetz, Edgar \& Chevalier 1988;
GB93; Paper II). Consistently with GTI theory (GB93), the density of the
shell oscillates with a period which is of the order of the cooling time
behind the shocks, $\tau \approx 2.2 t_{cool}$ (where $t_{cool} \approx
0.77 t_d$ for the HD1 model as obtained from the initial conditions,
or $\tau \approx 1.7t_d$). The introduction of the different magnetic
field configurations alters both the amplitude and the pattern of the
density oscillations, particularly in the presence of the toroidal
field (bottom panel, left), where the delay in the oscillations can be
interpreted as being due to the increase of the cooling distance behind
the shock caused by the amplification of the toroidal component (e.g.,
Paper I and II). In the right panels of Figure 4, we see that the density
of the IWSs also varies in time due to the global thermal instabilities,
with a period $\tau \approx 2t_d$, and the oscillation pattern is also more affected
in the presence of the toroidal field. In all cases, the oscillation peak
decays with time as a consequence of both the sideways deposition of the
shocked material into the cocoon and the damping of the GTI (GB93). [This
density decay is less obvious in the leading working surface (left panels)
because of the continuous accumulation of material in the head due to
the merging of the IWSs.]

\subsection{Effects of different pulsation periods}

Figure 5 displays the midiplane density contours (left) and the
velocity field distributions (right) for four pulsed jets after they
have propagated over a distance $\approx 40R_j$, at $t/t_d = 3$. The
top jet is purely hydrodynamic (model HD2 in Table 1); the second
(from top to bottom) has an initial constant longitudinal magnetic field
(model ML2 in Table 1); the third has an initial helical magnetic field
(MH2), and the bottom has an initial toroidal magnetic field (MT2). The
initial conditions are the same as in Figure 1, but here the period of
the injected sinusoidal velocity profile is twice as large ($P=1t_d$;
see Table 1). The four jets in Figure 5 are slightly advanced in the
computational domain when compared to those of Figure 1. This is a
natural consequence of the larger period $P$ of the injected velocity
variability (see, e.g., de Gouveia Dal Pino 2001).  As the average
propagation velocity of the leading bow shock ($v_{bs}$) is larger than
that of the jets of Figure 1, a smaller amount of shocked material is
accumulated behind the jet shock or Mach disk (whose velocity is $v_{Md}
\simeq v_j - v_{bs}$). Therefore, the dense, cold shell that develops at
the head due to the radiative cooling of the shocked material is thinner
and less clumpy in the larger period jets.

As before, the overall jet morphology is not very much affected by the
presence of the different magnetic field configurations, although the
tension forces associated with the toroidal magnetic field component in
the MHD jets with initially helical and toroidal geometries (third and
fourth panels) increases the jet collimation with respect to the HD jet.
Global thermal instabilities also occur in these models. As in Figure
4, Figure 6 shows that the leading working surface (top-left panel)
and the IWSs (here represented by the second IWS, top-right panel) of
the purely HD jet also undergo temporal density oscillations due to the
GTI. Likewise, the shape and amplitude of the oscillations are affected
by the introduction of the magnetic fields, particularly in the helical
(third panels) and toroidal (fourth panels) cases.

The H$\alpha$ intensity is known to be strong behind the knots of
the observed HH jets.  As discussed in Paper III, we can estimate the
H$\alpha$ emissivity ($I_{{\rm H}\alpha}$) behind the IWSs using the
results of our simulations and the relation given by Raga \& Cant\'o
(1998), $ I_{{\rm H}{\alpha}} \propto \rho_{d} v_s^{3.8} $, where
$\rho_{d}$ is the downstream pre-shock density, and $v_s$ is the shock
speed. In Paper III (see also de Gouveia Dal Pino \& Cerqueira 2001),
we have found for the MHD jets of Fig. 1 that the H$\alpha$ intensity
behind the internal knots is amplified up to factors $\sim$ 3, and 4  in
the presence of the helical and toroidal field geometries, respectively,
relative to the HD jet. The same evaluation perfomed for the jets of
longer period of Figure 5, results in amplifications of the H$\alpha$
intensity up to factors $\sim$ 5, and 4, for the MHD jets with helical and
toroidal fields, respectively, with respect to the H$\alpha$ intensity
in the HD jet, while for the MHD jet with initial longitudinal field,
the H$\alpha$ intensity is essentially the same as that of purely HD jet
(see  Figure 7).

\subsection{Effects of different values of $\beta$}

Figures 8 and 9 compare the early evolution of jets with different
magnetic field strengths. Figure 8 displays two MHD jets with initial
longitudinal magnetic field geometry: a $\beta \simeq 1$ (ML1, middle
panel), and a $\beta \simeq 0.1$ jet (ML1b, bottom panel), which are
compared with a HD jet (HD1, $\beta = \infty$, top panel). Likewise,
Figure 9 displays two MHD jets with initial helical magnetic field
geometries: a $\beta \simeq 1$ (MH1, middle panel), and a $\beta \simeq
0.1$ jet (MH1b, bottom), which are also compared with the HD jet (HD1,
top panel). The remaining initial conditions in these figures are the
same as in Figure 1. In the $\beta \simeq 0.1$ models, the maximum
initial magnetic field is $\approx 260 \mu$G. For both magnetic field
configurations, we note a considerable increase in the jet collimation
with larger $\bf{B}$-field ($\beta \simeq 0.1$) that is obviously caused
by the action of the much stronger confining magnetic tension and pressure
forces.  Nonetheless, the global characteristics of the flow found in
the HD, and $\beta \simeq 1$ jets are retained in the $\beta \simeq 0.1$
flows, particularly in the longitudinal field configuration (Figure 8).
For example, the clumpy structures that develop at the head of the HD,
and $\beta \simeq 1$ jets as a consequence of the combined effects of the
radiative cooling and Rayleigh-Taylor (R-T) instabilities (see Paper I
and II), are still present in the $\beta \simeq 0.1$ jet. This result is
somewhat in contradiction with the recent 2-D results of OR, who found
that the growth of the R-T instability and, therefore, the development
of fragmentation at the jet head, could be inhibited in the presence of
longitudinal fields even at equipartition conditions. This could be due
to the differences in the assumed input conditions of their model with
respect to the present ones. We note, however, that according to R-T
instability theory an enhancement of the instability is expected in the
presence of longitudinal $\bf{B}$-field (e.g., Jun, Norman \& Stone 1995).

Figure 10 depicts the velocity (top panel), density (second panel),
toroidal (third panel), and longitudinal (bottom panel) magnetic field
components along the beam axis \footnote{The toroidal magnetic field
component, $B_{\phi}$, has been actually taken close to the jet surface
other than along the jet's central axis where its value is too small
(see eq. 3b).} for the $\beta \simeq 1$ jet of Figure 9 (model MH1b in
Table 1), at $t/t_d = 2$.  As before, the toroidal component of the
magnetic field (third panel) sharpens within the knots and rarefies
between them, while the longitudinal component (forth panel) is stronger
between the knots. For the $\beta \simeq 1$ models with helical and
toroidal configurations (see Figs. 2b, and 9), we find that the toroidal
magnetic field components behind the IWSs are amplified by factors of
$\lesssim$ 1.5 with respect to the initial values. For the $\beta \simeq
0.1$ model (Figs. 9 and 10), we find only a slightly larger amplification
of $B_{\phi}$ behind the IWSs $\gtrsim 2$.

As in previous 2-D calculations (SH, OR), we also find that the
density behind the IWSs increases with increasing $\bf{B}$-strength
(or decreasing $\beta$), particularly in the presence of toroidal field
components (Figure 9). This density enhancement is caused again by the
confining tension forces associated with the toroidal fields which are
larger for lower $\beta$-jets. However, in our 3-D calculations, we find
smaller density amplifications. For example, for the $\beta \simeq 1$,
and 0.1 jets with helical fields (Figure 9), we find a maximum density
amplification behind the IWSs of a factor $\lesssim 1.4$, and $\lesssim
4$, respectively, relative to the HD jet, and for the $\beta \simeq 1$
jet with purely toroidal field (Fig. 1), we find a maximum density
amplification of a factor $\lesssim 1.6$. On the other hand, SH have
found larger density amplifications by factors $\sim 2-40$ in their 2-D
simulations of jets with purely toroidal fields and $\beta \simeq 1$, and
$\simeq 0.25$, respectively.  In the 2-D simulations of OR, the magnetic
compression of the IWSs by the toroidal fields is more drastic,
causing the disruption of them, even in $\beta \simeq 1$ jets. Although
these differences can be partially attributed to differences in the
input conditions of each model (see \S 1), the results above suggest
that the effects of the magnetic fields upon the jet structure tend to
be smoothed out in 3-D flows (see discussion below).

\section{Discussion and Conclusions}

We have presented the results of fully three-dimensional (3-D),
magnetohydrodynamic (MHD) numerical simulations of radiatively cooling,
overdense, pulsed jets, using the Smoothed Particle Magnetohydrodynamic
(SPMHD) technique.  We have explored the role of magnetic fields in pulsed
jets considering different magnetic field strengths ($\beta \simeq 0.1,
1$, and $\infty$) and configurations: longitudinal (aligned with the
jet axis), helical, and purely toroidal magnetic fields.  Our results
could be summarized as follows:

1 - Magnetic fields with intensities of a few 10$\mu$G, which are of the
order of those inferred for Herbig-Haro (HH) jets (e.g., Morse et al.
1992, 1994), and which are in close equipartition of energy
density with the thermal gas  are able to improve  collimation of the
jet with respect to a non-magnetic jet. Such collimation effect is,
as expected, more pronounced for magnetized jets with initial helical and
toroidal fields.

2 - Although of relevance for the formation of structures in the
shell at the jet head, the combined effect of non-uniform cooling and
the Rayleigh-Taylor (R-T) instability (e.g., GB93; Paper I), which is
particularly evident in the case of steady-state jets (see Paper I),
becomes less important in pulsed jets. In these cases,  the continuous
impact of the internal working surfaces (IWSs) with the jet head can also
cause shell fragmentation.  While the presence of toroidal magnetic
field components can modify the morphology close to the jet head inhibiting
its R-T fragmentation in  the early evolution of the jet, as previously
reported (paper I, III), the later impact of the pulsed-induced internal knots
causes the appearance of a clumpy, complex morphology at the jet head
(as required by the observations of Herbig-Haro jets) even in MHD 
models with helical or toroidal configurations.

3 - The overall morphology of the jet does not seem to be very much
affected by the presence of the different magnetic field configurations,
or different magnetic field strengths (see also paper III). Nonetheless,
the role of the magnetic fields in the determination of the detailed
structure and the emission properties behind the shocks in the head and
the IWSs cannot be ruled out.  The 3-D simulations show that the density
of the shell and IWSs undergoes oscillations with time (with periods of
the order of the cooling time behind the shocks) that are caused by global
thermal instabilities of the strong radiative shocks. The amplitude and
shape of these oscillations are affected by the introduction of magnetic
fields, particularly by toroidal configurations.

4 - The 3-D calculations also indicate that the the H$\alpha$ emissivity
can be enhanced by factors up to $\sim 4$, and 5 in the MHD models with
toroidal field components, and $\beta \simeq$ 1, and 0.1, respectively,
when compared with the non-magnetic calculation. Using the same data
from Hartigan, Raymond \& Hartman (1987; see also Hartigan, Morse \&
Raymond 1994) that have allowed to obtain $I_{{\rm H}\alpha} \propto
\rho_{d} v_s^{3.8}$ for the H$\alpha$ intensity (Raga \& Kofman 1992),
we find that the intensity of the [S II] doublet is wealky dependent on
the shock velocity ($I_{{\rm [S~II]}} \propto v_s^{0.2}$).  We can thus
infer from our results and from these two relations, that the $I_{{\rm
[S~II]}}/I_{{\rm H}\alpha}$ ratio (which is frequently used to determine
the excitation level of the  HH objects) can decrease by factors up
to $\sim 4-5$ in the MHD models with helical or  toroidal fields (for
$\beta \simeq 1-0.1$, respectively), with respect to the HD jet. Thus
if the HH jets are really imbedded in magnetic fields with strengths
not far from  the equipartition with the gas, the evaluations of the
physical quantities (such as, ionization fraction, mass loss ratio,
etc.), which are presently made without taking into account the magnetic
fields, could be significantly modified.  We note, however, that the
estimates above  of $I_{{\rm [S~II]}}/I_{{\rm H}\alpha}$ are very crude
and preliminary. Future calculations coupling the (magneto)hydrodynamic
calculation and the evolution of the chemical species will allow to
obtain more precise estimates of the intensity ratios above.  Also,
it's interesting to note that in the regime of IWS shock velocities, and
preshock densities and $\bf{B}$-fields examined here ($v_s \simeq 30$
km s$^{-1}$, $B < 300 \mu$G, $n_j \simeq 1000$ cm$^{-3}$), previous
``planar" shock calculations by Hartigan, Morse \& Raymond (1994) have
indicated that the $I_{{\rm [S~II]}}/I_{{\rm H}\alpha}$ ratio is less
sensitive to the magnetic field intensity than in our calculations.

5 - The differences that arise in 2-D simulations with different magnetic
field geometry and strengths seem to diminish in the 3-D flows. Like in
2-D models, the density behind the IWSs also increases with increasing
$\bf{B}$-strength (or decreasing $\beta$) in the 3-D calculations, but
at a lower rate.  For jets with initial helical field and  $\beta=1$,
and 0.1, we have found maximum density amplifications behind the IWSs of
factors $\lesssim 1.4$, and $\lesssim 4$ respectively, and for a $\beta
\simeq 1$ jet with pure toroidal field, we have found a maximum density
amplification of a factor $\lesssim 1.6$, relative to the HD jet. These
amplification factors are substantially smaller than those found in the
2-D models (SH; OR).

6- The amplification of the toroidal field components behind the IWSs
is also smaller in 3-D flows. We have found field amplification factors
of $\lesssim$ 1.5 for MHD models with $\beta \simeq 1$ and helical or
toroidal configurations, and amplification factors $\gtrsim 2$ in the
$\beta \simeq  0.1$ model with  helical configuration, with respect to
their initial magnetic fields.  These values are much smaller than
those detected, for instance, by SH, who found field amplifications
up to $\approx 7$ and 60 in their  $\beta \simeq 1$, and 0.25 models,
respectively.

7 - We have detected no signatures for the formation of nose cones. Such
elongated structures are often found to develop at the head of magnetized
jets in 2-D simulations involving toroidal magnetic fields. They are
formed as the shocked gas is confined between the Mach disk and the bow
shock by the tension forces associated with the toroidal field (e.g.,
Lind et al. 1989; K\"oss, M\"uller \& Hillebrandt 1990; OR; SH). Although
our 3-D calculations are probably limited by lower numerical resolution,
the typical extensions of the nose cones in the 2-D calculations are
few times the jet radius (SH; OR), so that they should be apparent even
in a low resolution calculation. Our results thus indicate that nose
cones are unstable in 3-D.  Perhaps more convincing is the fact that the
jet shock (or Mach disk) is found to be very close to the bow shock in
most of the HH jets, as is the case of the HH 47A which, following the
interpretation of Morse et al. (1994), may have a transverse magnetic
field of the order of $\approx 150 \mu$G.

8 - We speculate that the lower density amplification and the absence of
nose cones in the 3-D calculations could be, in part, explained by the
fact that, as the magnetic forces are intrinsically three-dimensional,
they cause part of  the material to be deflected in a third direction,
therefore, smoothing out the strong focusing of the shocked material
that is otherwise detected in the 2-D calculations (see also de Gouveia
Dal Pino \& Cerqueira 2001).

9 - Our simulations involving helical magnetic field geometry reveal the
development of a wiggling structure at the late evolution that is due to
the excitation of the kink mode of the Kelvin-Helmholtz instability. This
could potentially explain the gentle wandering often displayed by HH jets
(e.g., HH 111, e.g., Reipurth et al. 1997; HH 46/47, e.g., Heathcote et
al. 1996). The appearance of this mode whose development is not possible
in 2-D axisymmetric  models also reinforces the importance of fully
3-D calculations.

\acknowledgments

We are thankful to Alex Raga for his advice and suggestions.
A.H.C. would like to thank the Brazilian agency FAPESP, that have
supported  this work under a Ph.D. Fellowship program (process
96/08182-3). E.M.G.D.P. acknowledges the Brazilian agencies FAPESP and
CNPq for partial support. The simulations were performed on a cluster
of Linux-based PC's, whose purchase was made possible by FAPESP. Also,
partial support from the PRONEX/FINEP (41.96.0908.00) project is
acknowledged.

\clearpage

\figcaption[f1.ps]{Midiplane density contours (left) and the velocity
field distributions (right) for four supermagnetosonic, radiatively
cooling, pulsed jets after they have propagated over a distance $\approx
35R_j$, at $t/t_d = 3$ (where $t_d \equiv R_j/c_a \approx$ 38 years). The
top jet is purely hydrodynamic (model HD1 in Table 1); the second jet
(from top to bottom) has an initial constant longitudinal magnetic field
configuration (model ML1 in Table 1); the third jet has an initial helical
magnetic field configuration (ML1), and the bottom jet has an initial
toroidal magnetic field configuration (MT1). The initial parameters are
the same for all the models: $M_a=15$, $v_0=v_j[1+A\cdot{\rm sin}(2\pi
t / P)]$, with $v_j \simeq 250$ km s$^{-1}$, $A=0.25$ and $P=0.54t_d$
($\approx 20$ years), $\eta=5$, $\beta = \infty$ for the HD model, and
$\beta \simeq 1$ for MHD models.  The maximum density in each model (from
top to below) is: $n/n_a \approx$ 156, 173, 140 and 183 (1 $n_a = 200$
cm$^{-3}$). The $x$ and $z$ coordinates are in units of $R_j$. The jet
is injected into the computational domain at $x = -30R_j$. \label{fig1}}

\figcaption[f2.ps]{From top to bottom: velocity, density and
magnetic field component profiles along
the beam axis for the jets of Fig. 1 with: {\it a)} initial longitudinal
magnetic field (ML1), and {\it b)} initial toroidal field (MT1) . Time
displayed $t/t_d \simeq 3$.  The magnetic field intensities are displayed
in code units (one code unit is $\approx 21$ $\mu$G). Note that the
coordinate along the jet axis ($d$) has been shifted from $ -30 R_j$
to $0$.  (The noise we see near the head results from the approximation
of the first internal knot to the head.)  \label{fig2}}

\figcaption[f3.ps]{Midiplane density contours (top panel) and the
velocity field distributions (second panel) for: {\it a)} the HD1 model
(top panel of Fig. 1),  and {\it b)} the MH1 model (third panel of Figure
1), at $t/t_d =5$. The bottom panel of Fig. 2b shows the magnetic field
distribution.  \label{fig3}}

\figcaption[f4.ps]{Temporal evolution of the density in the leading
working surface (left) and for the second internal working surfaces
(right) for the four jets of Figure 1: HD1 (top), ML1 (second panel;
from top to bottom), MH1 (third panel) and MT1 (bottom).  \label{fig4}}

\figcaption[f5.ps]{The same as in Figure 1 but with
$P=1t_d$. \label{fig5}}

\figcaption[f6.ps]{The same as in Figure 4, but for the jets
of Figure 5. \label{fig6}}

\figcaption[f7.ps]{Ratio between the H$\alpha$ intensity along the jet
axis evaluated within the IWSs for the different magnetized jets of Figure
5 and the H$\alpha$ intensity of the purely hydrodynamical jet: MHD model
with longitudinal field (MHD L, squares), MHD model with helical field
(MHD H, bullets) and MHD model with toroidal field (MHD T, stars). The
$d$ coordinate along the jet axis has been shifted from $-30R_j$ to 0
(with respect to x  the coordinate of Figure 5). The symbols mark the
positions of the two prominent IWSs of the jets of Fig. 5. \label{fig7}}

\figcaption[f8.ps]{Midiplane density contours (left) and the velocity
field distributions (right) for three supermagnetosonic, radiatively
cooling, pulsed jets after they have propagated over a distance $\approx
20R_j$, at $t/t_d \simeq 1.75$. The top jet is purely hydrodynamical
(model HD1 in Table 1); the middle and bottom jets have an initial
constant longitudinal magnetic field configuration, with $\beta \simeq
1$, and $\beta \simeq 0.1$, respectively (models ML1, and ML1b of Table
1). The  $\beta = \infty$, and 1 models have been displayed also in
Figure 1 of Paper III).  \label{fig8}}

\figcaption[f9.ps]{Midiplane density contours (left) and the velocity
field distributions (right) for three supermagnetosonic, radiatively
cooling, pulsed jets after they have propagated over a distance $\approx
20R_j$, at $t/t_d \simeq 1.75$. The top jet is purely hydrodynamic
(model HD1 in Table 1); the middle and bottom jets have an initial
helical magnetic field configuration, with $\beta \simeq 1$, and $\beta
\simeq 0.1$, respectively (models MH1 and MH1b in Table 1; the $\beta =
\infty$, and 1 models have been  displayed also in Figure 1 of Paper III).
\label{fig9}}

\figcaption[f10.ps]{From top to bottom: velocity, density, toroidal
magnetic field component, and longitudinal magnetic field profiles along
the beam axis for the bottom-jet of Fig. 9 with initial helical magnetic
field. Time displayed $t/t_d \simeq 2 $. The magnetic field intensities
are displayed in code units (one code unit is $\approx 21$ $\mu$G).
The coordinate along the jet axis ($d$) has been shifted from $ -30 R_j$
to $0$. \label{fig10}}

\clearpage

\begin{deluxetable}{crrrrrr}
\tablecaption{Physical parameters of jets. \label{tbl-1}}
\tablewidth{0pt}
\tablehead{
\colhead{Model} & \colhead{$M_a$\tablenotemark{a}} & \colhead{$M_{{ms}_j}$}  &
\colhead{$M_{{ms}_a}$} & \colhead{$\eta$} & \colhead{$\beta$} &
\colhead{P}
          }
\startdata
HD1   & 15 & 33.5   & 15    & 5  & $\infty$ & 0.5  \\
ML1 & 15 & 22.6   & 10.1  & 5  & 1        & 0.5 \\
ML1b& 15 &  9.7   & 4.2   & 5  & 0.1      & 0.5 \\
MH1 & 15 & 23.2   & 10.4  & 5  & 1        & 0.5 \\
MH1b& 15 &  9.7   & 4.3   & 5  & 0.1      & 0.5 \\
MT1 & 15 & 26     & 11.6  & 5  & 1        & 0.5 \\
HD2   & 15 & 33.5   & 15    & 5  & $\infty$ & 1  \\
ML2 & 15 & 22.6   & 10.1  & 5  & 1        & 1 \\
MH2 & 15 & 23.2   & 10.4  & 5  & 1        & 1 \\
MT2 & 15 & 26     & 11.6  & 5  & 1        & 1 \\
\enddata

\tablenotetext{a}{This Mach number is evaluated for the average jet
speed $v_j=$ 250 km s$^{-1}$ (see equation 2).}

\end{deluxetable}

\end{document}